\documentclass[superscriptaddress,twocolumn,showpacs,prb]{revtex4}

\newcommand{\av}{_{\rm av}}

\usepackage{amsmath,graphicx,tabularx}
\usepackage{graphicx,amsmath}

\begin{document}

\title{
Study of the de Almeida-Thouless line using one-dimensional power-law diluted Heisenberg Spin Glasses}

\author{Auditya Sharma}
\affiliation{Department of Physics, University of California,
Santa Cruz, California 95064}
\author{A.~P.~Young}
\email{peter@physics.ucsc.edu}
\affiliation{Department of Physics, University of California,
Santa Cruz, California 95064}

\date{\today}

\begin{abstract}
We test for the presence or absence of the de Almeida-Thouless line using
one-dimensional power-law diluted Heisenberg spin glass model, in which the
rms strength of the interactions decays with distance, $r$ as
$1/r^{\sigma}$. It is argued that varying the power $\sigma$ is analogous to
varying the space dimension $d$ in a short-range model.
For $\sigma=0.6$, which is in the mean field regime regime, we find clear
evidence for an AT line. For $\sigma = 0.85$, which is in the non-mean-field
regime and corresponds to a space dimension of close to 3, we find no AT
line, though we cannot rule one out for very small fields. Finally for 
$\sigma=0.75$, which is in the non-mean-field regime but closer to the
mean-field boundary, the evidence suggests that there is an AT line, though
the possibility that even larger sizes are needed to see the asymptotic
behavior can not be ruled out.
\end{abstract}
%\pacs{03.67.Lx , 03.67.Ac, 64.70.Tg,75.10.Nr}
\maketitle

\section{Introduction}
\label{sec:intro}
One of the major unsolved questions in the field of spin glasses is the nature of
the low-temperature glassy phase. 
%Debate has raged for decades on whether or
%not the mean-field-theoretic picture is valid for realistic short-range
%models.
Two main candidate theories exist: the replica symmetry breaking (RSB) theory
which assumes that real spin glasses behave in a similar way to the
mean-field theory~\cite{parisi:80}, and the droplet
model~\cite{fisher:86,fisher:88}, which provides a phenomenological approach.
An important difference between the predictions of the two approaches is
whether or not there is a line of phase transitions, known as the 
de Almeida-Thouless~\cite{almeida:78} (AT) line, in the presence of a magnetic
field. An AT line is predicted in the RSB picture, but is argued not to exist
according to the droplet picture.
The existence of the AT line is arguably the most striking features of the immensely
complicated mean-field-theory of spin glasses~\cite{parisi:80}, and so it is
of intrinsic interest to know whether or not it occurs in real spin glasses. 
Some time ago Bray and Roberts~\cite{bray:80b} investigated whether or not there is an
AT line below six dimensions (six is the upper critical dimension for spin
glasses) using renormalization group ideas, but did not find a stable,
accessible fixed point. More recently, Temesvari~\cite{temesvari:08}
claimed to be able to
follow the AT line down to just below six dimensions.

Theoretically, it is interesting 
to understand spin glasses in a range of dimensions. However, this is hard
since one cannot simulate systems with a large number of spins $N$, where $N =
L^d$, for a \textit{range} of linear sizes $L$ (needed to do finite-size scaling) if
the dimension $d$ is large.
In particular, the
the mean-field regime, $d > 6$, is not directly amenable to
simulation.
Hence, instead,
we investigate a \textit{one-dimensional} model in which the rms strength of
the interactions fall off with a power $\sigma$ of the distance.
%power-law model with
%Heisenberg spins.
%larger system sizes than before,
This model has the advantage of allowing one to study large (linear) sizes, 
Furthermore, it has been suggested that changing the value of $\sigma$ is analogous to
changing the value of the space dimension $d$ in a short-range model. Consequently we
can study models in both the mean-field and non-mean-field regimes. The nature
of the spin glass phase in long-range models was discussed recently by
Moore~\cite{moore:10}.

Most previous work~\cite{leuzzi:08,leuzzi:09,young:04,katzgraber:05,katzgraber:09}
on one-dimensional, long-range spin glass models, used Ising spins.
However, in this paper, we
build on a recent work by us~\cite{sharma:10} which showed that, in mean field
theory, an AT line also occurs in $m$-component vector spins provided the
magnetic field is random in direction. More precisely, we perform Monte Carlo
simulations on the
three-component (Heisenberg) spin glass in one dimension with long-range
interactions in the presence of a random magnetic field.
This work follows 
on from a recent paper~\cite{sharma:11a} where we considered the same model in
zero-field.

The plan of this paper is as follows: in Sec.~\ref{sec:model} we describe the model and
the Monte Carlo method used to simulate it, in Sec.~\ref{sec:results} we describe the
results, and finally in Sec.~\ref{sec:conclusions} we summarize our conclusions. 

\section{Model and Method}
\label{sec:model}
The Hamiltonian we study is
\begin{equation}
\mathcal{H} = -\sum_{\langle i, j \rangle} J_{ij} \mathbf{S}_i \cdot \mathbf{S}_j -
\sum_i \mathbf{h}_i \cdot \mathbf{S}_i \, ,
\label{Ham}
\end{equation}
where the $\mathbf{S}_i$, $i = 1, 2, \cdots, N$,
are classical $3$-component Heisenberg spins of unit length,
and the interactions
$J_{ij}$ are independent random variables with zero mean and a variance 
which falls off with a power of the distance $r_{ij}$ between the
spins,
\begin{equation}
[J_{ij}^{2}]_{av} \propto \frac{1}{r_{ij}^{2\sigma}}. 
\label{J2}
\end{equation} 
The notation $[\cdots]\av$ indicates an average over the quenched disorder.
In addition we set $J_{ii} = 0$.
%Furthermore, following Ref.~[\onlinecite{sharma:11a}],
The magnetic fields $h_i^\mu$, where $\mu$ denotes a cartesian spin component,
are chosen to be
independent Gaussian random fields, uncorrelated between
sites, with zero mean, which satisfy
\begin{equation}
[ h_i^\mu h_j^\nu]\av = h_r^2\, \delta_{ij}\, \delta_{\mu\nu} \, .
\label{hs}
\end{equation}

Following Leuzzi et.~al.~\cite{leuzzi:08}, and continuing along the lines of
the zero-field paper~\cite{sharma:11a}, the interactions of our model are such
that, instead of the \textit{magnitude} of the interaction falling off with
distance like Eq.~\eqref{J2}, it is the \emph{probability} of there being a
non-zero interaction between sites $(i,j)$ which falls off, and when an
interaction does occur, its variance is independent of $r_{ij}$. The mean
number of non-zero interactions from a site, which we call $z$, can be fixed,
and here we take $z = 6$. To generate the set of pairs $(i,j)$ that have an
interaction with the desired probability
%is converted into: \begin{equation} P[J_{ij} \neq 0] \propto
%\frac{1}{r_{ij}^{2\sigma}}.  \end{equation} This has the advantage of making
%the energy $O(N)$ as opposed to $O(N^2)$, and therefore allows the simulation
%of larger size lattices. To generate bonds satisfying this condition,
%we start by fixing the total number of bonds in the $N$ site 
%system as $Nb=N*z/2$, where $z \equiv 6$ is the average coordination number. In contrast to Leuzzi et al~\cite{leuzzi:08}, 
%we prefer using the geometric distance between sites $i,j$ is defined as:
%\begin{equation}
%r_{ij}=\frac{N}{\pi}sin[\frac{\pi}{N}(i-j)].
%\end{equation}
%Following Leuzzi et al~\cite{leuzzi:08}, we produce these bonds as follows. 
we choose spin $i$ randomly, and then choose
$j \ (\ne i)$ at distance $r_{ij}$ with probability 
\begin{equation}
p_{ij} = \frac{r_{ij}^{-2\sigma}}{\sum_{j\, (j\neq i)}r_{ij}^{-2\sigma}} \, ,
\end{equation}
where, for $r_{ij}$, we put the sites on a circle and use the distance of the
chord, i.e.
\begin{equation}
r_{ij}=\frac{N}{\pi}\sin\left[\frac{\pi}{N}(i-j)\right].
\end{equation}
If $i$ and $j$ are already connected, we repeat the process until we find a pair
which has not been connected before. We then connect $i$ and $j$ with an
interaction picked from a Gaussian interaction
whose mean is zero and whose standard deviation is $J$, which we set equal to 1.
This process is
repeated precisely $N_b = z N / 2 $ times. 

The result is that each pair $(i, j)$ will be
connected with a probability $P_{ij}$ which must satisfy the condition
$N \sum_j P_{ij} = N z$ since $P_{ij}$ only depends on $|i-j|$, $P_{ii}=0$, and
there are precisely $N z /2 $ connected pairs. 
It follows that, for a fixed site $i$,
\begin{equation}
\sum_j [\,J_{ij}^{2}\,]_{av} = J^2 \sum_j P_{ij} = J^2 z \, .
\label{sumJij2}
\end{equation}
The mean-field spin glass
transition temperature for $m$-component vector spins is given, for zero
field,
by~\cite{almeida:78b}
\begin{equation}
T_c^{MF} = \frac{1}{m} \left(\sum_j [\,J_{ij}^{2}\,]_{av}\right)^{1/2} 
= {\sqrt{z} \over m} \, J \, ,
\label{TcMF}
\end{equation}
where the last equality follows from Eq.~\eqref{sumJij2}.
In mean-field theory, the critical magnetic field at zero-temperature for
$m$-component vector spins~\cite{sharma:10} is 
(after accounting for the
different normalization of the spins in
Ref.~[\onlinecite{sharma:10}]) 
given by
\begin{align}
h_{c}^{MF} \equiv h^{MF}_{AT}(T=0) &= {\sqrt{\frac{m}{m-2}}}\,\, T_{c}^{MF}
\nonumber \\
&= \sqrt{\frac{z}{m(m-2)}}\,\, J, 
\end{align}

We set $J=1$ so that, for the situation here,
\begin{equation}
J = 1,\ z=6,\ m=3\, ,
\end{equation}
we have
\begin{equation}
T_c^{MF} = {\sqrt{6} \over 3} \simeq 0.816\, ,
\end{equation}
the same as for the nearest-neighbor Heisenberg spin glass on a simple cubic
lattice, and
\begin{equation}
h_c^{MF} = \sqrt{2} \simeq 1.414.
\end{equation}
The ratio of these two quantities, which we shall refer to later, is
given by
\begin{equation}
{h_c^{MF} \over T_c^{MF}} = \sqrt{3} \simeq 1.732.
\label{hcMF}
\end{equation}

According to Ref.~[\onlinecite{sharma:10}], a good approximation for the AT
line in the mean field theory of the Heisenberg ($m=3$) spin glass is
\begin{equation}
{h^{MF}_{AT}(T) \over T_c^{MF}} = \left({4 m \over m + 2}\, t^3\right)^{1/2} \, 
\label{hoverTc}
\end{equation}
where 
\begin{equation}
t = {T - T_c^{MF} \over T_c^{MF}} \, .
\label{t}
\end{equation}
In Eq.~\eqref{hoverTc} we have again allowed
for the different normalization of the spins in Ref.~[\onlinecite{sharma:10}].
Equation \eqref{hoverTc} is exact, in mean field theory, near $T_c^{MF}$, and, for $m=3$,
works very well down to quite low temperatures, see Fig.~1 in
Ref.~[\onlinecite{sharma:10}]. Even for $T = 0\, ,(t = 1)$, Eq.~\eqref{hoverTc}
gives $h_c^{MF} / T_c^{MF} = \sqrt{12/5} \simeq 1.549$, whereas the correct value in
mean field theory is, according to Eq.~\eqref{hcMF}, $h_c^{MF} / T_c^{MF}
\simeq 1.732$.

We perform Monte Carlo simulations for this model in a magnetic field for three 
values of $\sigma$: $0.6$, $0.75$, and $0.85$. As discussed in the zero-field 
paper~\cite{sharma:11a}, $0.6$ lies in the mean-field regime,
%where an AT line would be expected a priori,
while the other two values of $\sigma$
are in the non-mean-field regime.
%where the question of the AT line is open.
%The value
%$\sigma = 0.85$ is expected to correspond  fairly closes to realistic $3-d$
%nearest-neighbor spin glasses.

An approximate connection between a value of $\sigma$ and the effective
dimension of an equivalent short-range model, $d_{\rm eff}$,
is~\cite{larson:10}
\begin{equation}
d_{\rm eff} = {2 \over 2 \sigma - 1} \, .
\label{deff}
\end{equation}
In the non mean-field region, a more accurate connection, which involves the
exponent $\eta_{SR}$ of the short-range model, can also be
obtained~\cite{larson:10}, but we will neglect this correction here since we
have little information on $\eta_{SR}$. From Eq.~\eqref{deff} the effective
dimensions corresponding to $\sigma = 0.75$ and 0.85 are $d_{\rm eff} =4$ and
$d_{\rm eff} \simeq 3$, respectively.

We continue to use the technology described in the zero-field
paper~\cite{sharma:11a}: overrelaxation sweeps, heatbath sweeps, and parallel tempering.
We perform one heatbath sweep, and one parallel tempering sweep for every ten overrelaxation
sweeps.

The Gaussian nature of the interactions and the magnetic fields affords
a useful test
for equilibration.
The relation
\begin{equation}
U = {J^2 \over T} \, {z \over 2} \, (q_l - q_s) \, + {h_{r}^{2} \over T} \, (q-1),
\label{equiltest}
\end{equation}
is valid in equilibrium but, very plausibly, the two sides approach
their common equilibrium
value from opposite directions as equilibrium is approached. Here
\begin{equation}
U = -{1 \over N}
\Bigl[\, \sum_{\langle i, j \rangle} \epsilon_{ij} J_{ij} \langle {\bf S}_i \cdot
{\bf S}_j \rangle + \sum_{i, \mu} h_i^\mu 
\langle S_i^\mu \rangle \, \Bigr]\av 
\label{Uav}
\end{equation}
is
the average energy per spin,
$q = (1/N)\sum_{i}[\langle{\bf S}_{i}\rangle \cdot \langle{\bf S}_{i}\rangle]\av$ 
is the Edwards-Anderson order parameter,
$q_l = (1/N_{b})\sum_{\langle i, j \rangle} \epsilon_{ij} 
[ \langle 
{\bf S}_i \cdot {\bf S}_j \rangle^2]\av$ is the ``link overlap'', and
$q_s = (1/N_{b})\sum_{\langle i, j \rangle} \epsilon_{ij}
[\langle ({\bf S}_i \cdot {\bf S}_j)^2
\rangle]\av$, where $N_{b}=z N/2$,
and $\epsilon_{ij} = 1$ if there is a bond between $i$ and $j$ and is zero
otherwise.
Equation~(\ref{equiltest}) is easily derived by integrating by parts
Eq.~\eqref{Uav}
with respect to $J_{ij}$ and $h_{i}^\mu$ since they have
Gaussian distributions.

We determine both sides of
Eq.~\eqref{equiltest} for different numbers
of Monte Carlo sweeps (MCS) which increase
in a logarithmic manner, each value being twice the previous one. In all
cases we
average over the last half of the sweeps. We consider the data to be
equilibrated, if, when averaging over a large number of samples,
Eq.~\eqref{equiltest} is satisfied for at least the last two points.
Note that in the numerics we set $J = 1$.
Table \ref{simparams} lists the parameters of the simulation.

\begin{table}
\caption{
Parameters of the simulations.  $N_{\rm samp}$ is the number of samples,
$N_{\rm equil}$ is the number of overrelaxation Monte Carlo sweeps for
equilibration for each of the $2 N_T$ replicas for a single sample. The same
number of sweeps is done in the measurement phase, with a measurement
performed every four overrelaxation sweeps.
%$N_{\rm meas}$ is the number of overrelaxation sweeps for measurement.
The number of heatbath sweeps is equal to
10\% of the number of
overrelaxation sweeps.  $T_{\rm min}$ and $T_{\rm max}$ are the
lowest and highest temperatures simulated,
and $N_T$ is the number of temperatures
used in the parallel tempering.
\label{simparams}
}
\begin{tabular*}{\columnwidth}{@{\extracolsep{\fill}}r r r r r r r r r}
\hline
\hline
$\sigma$ & $h_{r}$ & $N$  & $N_{\rm samp} $ & $N_{\rm equil}$ & $T_{\rm min}$ &
$T_{\rm max}$ & $N_{T}$  \\
\hline
$0.6$ & $0.1$ &  $128$ &  $6000$ &   $512$ & $0.20$ & $0.50$ &  $21$ \\
$0.6$ & $0.1$ &  $256$ &  $6000$ &   $512$ & $0.20$ & $0.50$ &  $23$ \\
$0.6$ & $0.1$ &  $512$ &  $6000$ &  $2048$ & $0.20$ & $0.50$ &  $24$ \\
$0.6$ & $0.1$ & $1024$ &  $6000$ &  $4096$ & $0.20$ & $0.50$ &  $27$ \\
$0.6$ & $0.1$ & $2048$ &  $6000$ &  $8192$ & $0.20$ & $0.50$ &  $30$ \\
$0.6$ & $0.1$ & $4096$ &  $2000$ & $16384$ & $0.25$ & $0.45$ &  $40$ \\
$0.6$ & $0.1$ & $8192$ &  $1000$ & $32768$ & $0.27$ & $0.45$ &  $42$ \\
$0.6$ & $0.1$ &$16384$ &  $1000$ & $16384$ & $0.30$ & $0.42$ &  $45$ \\
\hline
$0.75$ & $0.1$ &  $128$ &  $6000$ &   $8192$ & $0.05$ & $0.24$ &  $21$ \\
$0.75$ & $0.1$ &  $256$ &  $6000$ &  $16384$ & $0.05$ & $0.24$ &  $22$ \\
$0.75$ & $0.1$ &  $512$ &  $6000$ &  $16384$ & $0.10$ & $0.30$ &  $24$ \\
$0.75$ & $0.1$ & $1024$ &  $3500$ &  $65536$ & $0.10$ & $0.24$ &  $20$ \\
$0.75$ & $0.1$ & $2048$ &  $2300$ & $131072$ & $0.10$ & $0.24$ &  $23$ \\
$0.75$ & $0.1$ & $4096$ &  $1000$ & $262144$ & $0.10$ & $0.30$ &  $45$ \\
\hline
$0.85$ & $0.05$ &  $128$ &  $10000$ &  $32768$ & $0.015$ & $0.20$ &  $25$ \\
$0.85$ & $0.05$ &  $256$ &  $9000$  &  $65536$ & $0.015$ & $0.20$ &  $34$ \\
$0.85$ & $0.05$ &  $512$ &  $5000$  & $262144$ & $0.02$  & $0.20$ &  $38$ \\
$0.85$ & $0.05$ & $1024$ &  $5000$  & $524288$ & $0.03$  & $0.20$ &  $40$ \\
$0.85$ & $0.05$ & $2048$ &  $4100$  & $524288$ & $0.05$  & $0.20$ &  $34$ \\
\hline
$0.85$ & $0.1$ &  $128$ &  $6000$ &    $8192$ & $0.05$ & $0.25$ &  $21$ \\
$0.85$ & $0.1$ &  $256$ &  $2500$ &   $16384$ & $0.05$ & $0.25$ &  $22$ \\
$0.85$ & $0.1$ &  $512$ &  $2000$ &   $32768$ & $0.05$ & $0.25$ &  $24$ \\
$0.85$ & $0.1$ & $1024$ &  $1200$ &  $262144$ & $0.05$ & $0.20$ &  $34$ \\
$0.85$ & $0.1$ & $2048$ &  $1600$ &  $262144$ & $0.05$ & $0.20$ &  $40$ \\
$0.85$ & $0.1$ & $4096$ &  $1800$ &  $131072$ & $0.10$ & $0.20$ &  $30$ \\
\hline
\hline
\end{tabular*}
\end{table}

%The main physical quantities that are measured in our simulation are the spin glass
%susceptibility $\chi_{SG}$ at wave vectors $k=0$ and ${k}_\mathrm{min} = (2\pi/N)$, from which
%the spin glass correlation length $\xi_{SG}$ is then obtained. 

%The spin glass order parameter at wavevector $k$ is given by
%\begin{equation}
%q^{\mu\nu}({k}) = {1 \over N} \sum_{i=1}^N S_i^{\mu(1)} S_i^{\nu(2)}
%e^{i {k} \cdot {R}_i},
%\end{equation}
%where $\mu$ and $\nu$ are spin components, and ``$(1)$'' and ``$(2)$''
%denote two
%identical copies of the system with the same interactions. From this we
%determine the wave vector dependent
%spin glass susceptibility $\chi_{SG}({k})$ by
%\begin{equation}
%\chi_{SG}({k}) = N \sum_{\mu,\nu} [\langle \left|q^{\mu\nu}({k})\right|^2 \rangle ]\av ,
%\end{equation}
%where $\langle \cdots \rangle$ denotes a thermal average and
%$[\cdots ]\av$ denotes an average over disorder.

%To determine the existence of an AT line,
%we compute the two-point finite-size correlation
%length \cite{palassini:99b,ballesteros:00,young:04}. For this we start
We determine the wave-vector-dependent spin-glass susceptibility,
given by~\cite{sharma:10}
%\begin{equation}
%\chi_{\rm SG}(k) = \frac{1}{N} \sum_{i, j} \left[\Big(
%\langle S_i S_j\rangle_T - \langle S_i \rangle_T \langle S_j\rangle_T
%\Big)^2 \right]_{\rm av}\!\!\!\!\! e^{ik\, (i-j)} ,
%\label{eq:chisg}
%\end{equation}
\begin{subequations}
\label{eq:chisg}
\begin{equation}
\chi_{\rm SG}(k) = {1 \over N}
\sum_{i, j} {1 \over m}
\sum_{\mu,\nu}
\Bigl[
(\chi_{i j}^{\mu\nu})^2 \Bigr]\av e^{ik\, (i-j)}, 
\end{equation}
where
\begin{equation}
\chi_{i j}^{\mu\nu} = 
\langle S_i^\mu S_j^\nu \rangle - \langle
S_i^\mu \rangle \langle S_j^\nu \rangle ,
\end{equation}
\end{subequations}
in which $\langle \cdots \rangle$ denotes a thermal average and
$[\cdots]_{\rm av}$ an average over the disorder. To avoid bias, each
thermal average is obtained from a separate copy of the spins, so we
simulate four copies at each temperature.
The spin glass correlation
length is then determined from
\begin{equation}
\xi_{SG} = {1 \over 2 \sin (k_\mathrm{min}/2)}
\left({\chi_{SG}(0) \over \chi_{SG}({k}_\mathrm{min})} - 1\right)^{1/(2\sigma-1)},
\end{equation}
where ${k}_\mathrm{min} = (2\pi/N)$.

According to finite-size
scaling~\cite{fss:gtlcd,sharma:11a}, the correlation length of the finite-system varies,
near the transition temperature $T_c$,
as
\begin{subequations}
\label{eq:xiscale}
\begin{align}
\label{xi_nonmf}
{\xi \over N} &= {\mathcal X} [ N^{1/\nu} (T - T_c) ]
%&& {\xi}/{N} &\sim {\mathcal X} [ N^{1/\nu} (T - T_c) ]
\; , \ (2/3 \le \sigma < 1), 
\\
{\xi \over N^{\nu/3}} &= {\mathcal X} [ N^{1/3} (T - T_c) ]
%&& {\xi_N}/{N^{\nu/3}} &\sim {\mathcal X} [ N^{1/3} (T - T_c) ]
\;,\ (1/2 <\sigma \le 2/3),
\label{xi_mf}
\end{align}
\end{subequations}
in which $\nu$, the correlation length exponent, is given, in the mean-field
regime, by $\nu = 1/(2\sigma - 1)$.
%We will use Eq.~\eqref{eq:xiscale}
%for both the spin glass correlation length $\xi_{SG}$, in which $T_c$
%will be set to $T_{SG}$, and the chiral glass correlation length $\xi_{CG}$,
%in which $T_c$ will be set to $T_{CG}$.
It follows that,
if there is a transition at $T = T_c$,
data for $ {\xi}/{N}$ ($ {\xi}/{N^{\nu/3}}$ in the mean-field region)
for different system sizes $N$ should cross at $T_c$.

We also present data for $\chi_{SG} \equiv \chi_{\rm SG}(0)$, which
has the finite-size scaling form
\begin{subequations}
\label{eq:chisgscale}
\begin{align}
\label{chi_nonmf}
{\chi_{\rm SG} \over N^{2 -\eta}} &= {\mathcal C}[N^{1/\nu} (T - T_c)]
\; , \ (2/3 \le \sigma < 1), 
\\
{\chi_{\rm SG} \over N^{1/3}} &= {\mathcal C}[N^{1/3} (T - T_c)]
\; , \ (1/2 < \sigma \le 2/3).
\label{chi_mf}
\end{align}
\end{subequations}
Hence curves of $\chi_{\rm SG}/N^{2 - \eta}$ ($\chi_{\rm SG} /
N^{1/3}$ in the mean-field regime) should also intersect. This is particularly
useful for long-range models since $\eta$ is given by the simple expression $2
- \eta = 2 \sigma - 1$ {\it exactly}.

In practice, there are corrections to this finite-size-scaling, so
data for different sizes do not all intersect at the exactly the same
temperature.
Including leading corrections to scaling, the intersection
temperature $T^{*}(N,2N)$ for sizes $N$ and $2N$ varies
as~\cite{binder:81b,ballesteros:96a,hasenbusch:08b,larson:10}
\begin{equation}
T^{*}(N,2N) = T_{c} + \frac{A}{N^{\lambda}},
\label{Tstar}
\end{equation}
where $A$ is the amplitude of the leading correction, and, in the non
mean-field regime, the
exponent $\lambda$ is given by
\begin{equation}
\lambda = \frac{1}{\nu}+\omega
\label{lambda}
\end{equation}
where $\omega$ is the leading correction to scaling
exponent.

\section{Results and Analysis}
\label{sec:results}

\begin{figure}
\begin{center}
\includegraphics[width=\columnwidth]{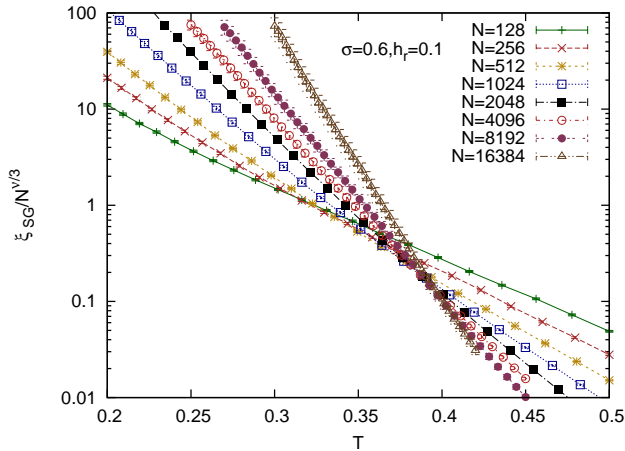}
\caption{(Color online)
A finite size scaling plot of data for
$\xi_{SG}$, for $\sigma=0.6$ in
a magnetic field of $h_{r}=0.1$. The intersections indicate that there
is a spin glass phase transition (AT line) in the presence of a magnetic field.
}
\label{fig:corrsg_0.6}
\end{center}
\end{figure}

\begin{figure}
\begin{center}
\includegraphics[width=\columnwidth]{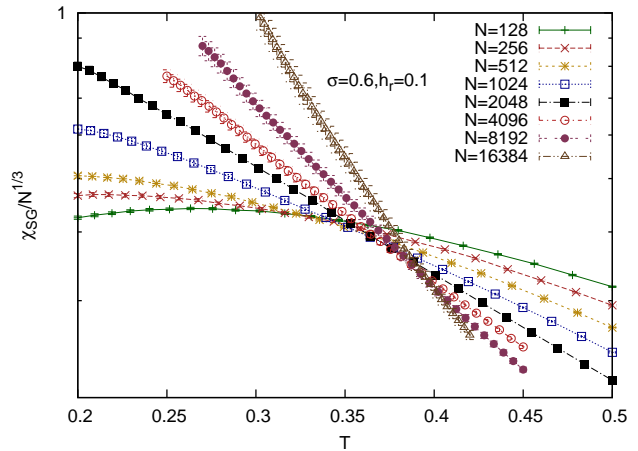}
\caption{(Color online)
A finite size scaling figure of data for $\chi_{SG}$, for $\sigma=0.6$ with
$h_{r}=0.1$. As in Fig.~\ref{fig:corrsg_0.6},
the intersections indicate the presence of an
AT line.
}
\label{fig:khisg_0.6}
\end{center}
\end{figure}

\begin{figure}
\begin{center}
\includegraphics[width=\columnwidth]{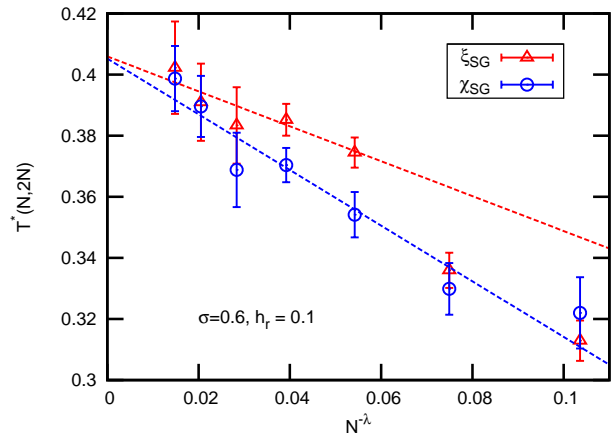}
\caption{(Color online)
A plot of the intersection temperatures
$T^{*}(N,2N)$ for $\sigma = 0.6$, obtained from the data in
Figs.~\ref{fig:corrsg_0.6} and \ref{fig:khisg_0.6},
as a function of
$N^{-\lambda}$, with $\lambda$
fixed to its value $0.467$, which is known exactly because $\sigma=0.6$ is
in the mean-field regime. The fits give: $T_{SG}=0.406 \pm 0.012$
from $\xi_{SG}$ and $T_{SG} = 0.405 \pm 0.007$ from $\chi_{SG}$. For
comparison, the zero field transition temperature
is~\cite{sharma:11a} $T_{SG} \simeq 0.563$.
}
\label{fig:fit_0.6}
\end{center}
\end{figure}

\subsection{$\sigma=0.6$}

We recall that $\sigma=0.6$ lies in the mean-field regime. In this regime,
simulations of the corresponding Ising model~\cite{katzgraber:09,leuzzi:09}
found an AT line.
%This means that one
%would expect mean-field theoretic ideas would hold for this model, which in
%turn means that an AT line would be expected for this case. In other words,
%the application of a small magnetic field at low enough temperatures, does not
%destroy the spin glass order.
From our plots for the Heisenberg spin glass
in Figs.~\ref{fig:corrsg_0.6} and \ref{fig:khisg_0.6} for
$h_r = 0.1$,
we come to the same
conclusion here.
According to Eqs.~\eqref{xi_mf} and \eqref{chi_mf},
data for $\xi_{SG} / N^{\nu/3}$
and $\chi_{SG} / N^{1/3}$ should intersect at the transition temperature.
We do, indeed find intersections, though the intersection temperatures vary
somewhat with size. 

The intersection temperatures are shown in Fig.~\ref{fig:fit_0.6}.
Fitting the intersection temperatures to Eq.~\eqref{Tstar}, using the known
value~\cite{larson:10,sharma:11a} $\lambda = 0.467$ we find
$T_{SG} = 0.406 \pm 0.012$ from $\xi_{SG}$ (omitting the two smallest sizes), and
$T_{SG} = 0.405 \pm 0.007$ from $\chi_{SG}$ (including all the data).
These two results agree well with each other.
%While these two results differ by
%slighlty more than would be expected from the errors, the difference is
%plausibly explained by small systematic effects for this range of sizes. Note
Note that the intersection temperatures \textit{increase} with increasing size,
which suggests that they will not disappear in the thermodynamic limit.

It is interesting to compare this point on the AT line, $(T, h_{AT}(T)) =
(0.405, 0.1)$,
with mean field predictions.
Replacing $T_c^{MF}$ with the actual zero field transition temperature
of~\cite{sharma:11a}
$T_c = 0.563$ in
Eqs.~\eqref{hoverTc} and $\eqref{t}$, 
we find that $T = 0.405$ gives a field $h_{AT}(T=0.405) = 0.130$,
which is slightly
larger than the actual field value of $0.1$. Hence the value of the field on the
AT line for $\sigma = 0.6$ is \textit{somewhat} less than that expected in mean field
theory, even allowing for the reduction in $T_c$ from its mean field value.

To conclude this section, the data suggests that there is an AT line for
$\sigma = 0.6$, as found earlier~\cite{katzgraber:09,leuzzi:09}
for the corresponding Ising model.

\begin{figure}
\begin{center}
\includegraphics[width=\columnwidth]{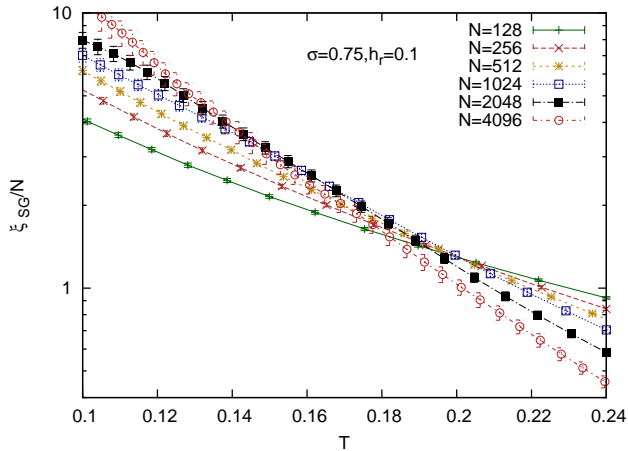}
\caption{(Color online)
A finite size scaling plot of data for $\xi_{SG}$ for $\sigma=0.75$. The magnetic
field is $h_{r}=0.1$.
%This figure shows that there is a spin glass phase transition in the presence
%of a small magnetic field. 
}
\label{fig:corrsg_0.75}
\end{center}
\end{figure}

\begin{figure}
\begin{center}
\includegraphics[width=\columnwidth]{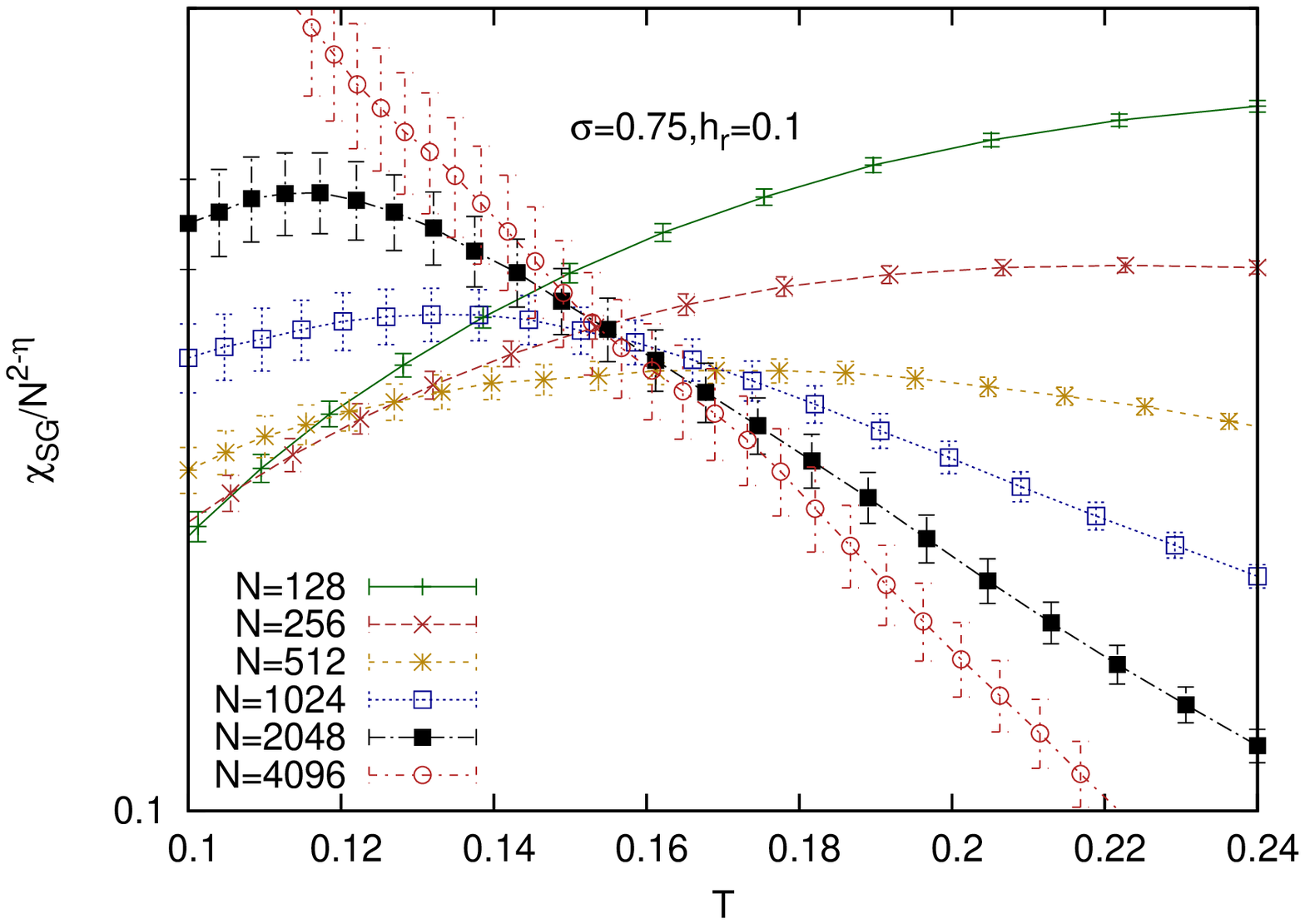}
\caption{(Color online)
A finite size scaling plot of data for $\chi_{SG}$ for $\sigma=0.75$.
The magnetic field is $h_{r}=0.1$.
%This figure shows that there
%is a spin glass phase transition in the presence of a small magnetic field.
}
\label{fig:khisg_0.75}
\end{center}
\end{figure}

\begin{figure}
\begin{center}
\includegraphics[width=\columnwidth]{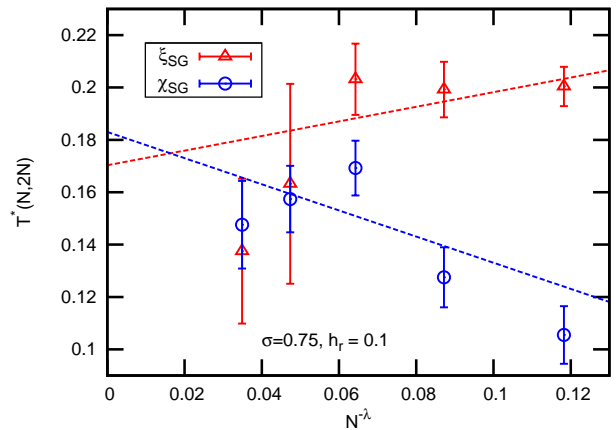}
\caption{(Color online)
A plot of the intersection temperatures
$T^{*}(N,2N)$ for $\sigma = 0.75$, obtained from the data in
Figs.~\ref{fig:corrsg_0.75} and \ref{fig:khisg_0.75}, as a function of
$N^{-\lambda}$ with $\lambda = 0.44$, the value found for $\sigma = 0.75$
in our \textit{zero field} study~\cite{sharma:11a}.
We were not able to determine $\lambda$ with any degree of precision
from the data in Figs.~\ref{fig:corrsg_0.75} and \ref{fig:khisg_0.75}.
Both sets of data are compatible with a transition
temperature of about 0.16 though there is also a hint in the data that the
intersection temperatures start to drop for the largest sizes.
For comparison, the zero field transition temperature is~\cite{sharma:11a}
$T_c \simeq 0.357$.
%The
%estimates for the transition temperatures are $T_{SG} = 0.170 \pm 0.022$ from
%$\xi_{SG}$ and $T_{SG} = 0.196 \pm 0.015$ from $\chi_{SG}$. For comparison,
%the corresponding zero-field transition temperature is about $0.357$.
%%$T_{SG} = 0.359 \pm 0.003$
%%from $\xi_{SG}$ and $T_{SG} = 0.354 \pm 0.005$ from $\chi_{SG}$.
}
\label{fig:fit_0.75}
\end{center}
\end{figure}

\subsection{$\sigma=0.75$}
According to Eq.~~\eqref{deff} (which is approximate) this value of $\sigma$
corresponds to a short-range model in $d = 4$ dimensions.
For the corresponding Ising study,
Ref.~[\onlinecite{katzgraber:09}]
did not find an AT line, though this conclusion was subsequently challenged in
Ref.~[\onlinecite{leuzzi:09}].

Data for the spin glass correlation length and susceptibility are shown
in Figs.~\ref{fig:corrsg_0.75} and \ref{fig:khisg_0.75}. There are clearly
intersections.
Since we are now in the non mean-field regime, the
exponent $\lambda$ in Eq.~\eqref{Tstar} is not known and so should be treated
as a fit parameter.
However, the intersection temperatures have quite large error
bars, and do not seem to vary monotonically, as shown in
Fig.~\ref{fig:fit_0.75}.
Thus the data is not of good enough quality to determine $\lambda$, and, in
the plot, we have, rather arbitrarily used the value $\lambda = 0.44$ obtained for
$\sigma = 0.75$ in our zero-field study~\cite{sharma:11a}.

Both sets of data are compatible with a non-zero $T_c$ of about 0.16, which is
to be compared with the zero field spin glass transition temperature $T_{SG}
\simeq 0.357$.
However, there is a suggestion in the data that the values of
$T^\star$ decrease for the largest pairs of sizes, so perhaps one should treat
this conclusion with some caution.

As we did above for $\sigma = 0.6$, we compare the putative point on the AT 
line, $(T, h_{AT}(T)) = (0.16, 0.1)$,
with mean field predictions.
Replacing $T_c^{MF}$ with the actual zero field transition temperature
of~\cite{sharma:11a}
$T_c = 0.357$ in
Eqs.~\eqref{hoverTc} and $\eqref{t}$,
we find that $T = 0.16$ gives a field $h_{AT}(T=0.16) = 0.227$,
which is considerably
larger than the actual field value of $0.1$.
Hence, if the intersections in
Figs.~\ref{fig:corrsg_0.75} and \ref{fig:khisg_0.75}, do represent a
transition in a field, the value of this AT field is
is \textit{considerably} less than that expected in
mean field
theory, even allowing for the reduction in $T_c$ from its mean field value.

%The Ising version of this model for $\sigma=0.75$ showed contrasting behavior
%in two different studies with the same model~\cite{leuzzi:09,katzgraber:09}.
%Our results seem to suggest that there is likely an AT line for this case. The
%reason for the ambiguity is that the two different quantities considered for
%finite size scaling, $\xi_{SG}$ and $\chi_{SG}$ appear to converge from
%different sides as the size is cranked up to the thermodynamic limit. Also,
%our data has rather big error bars for the intersection temperatures for the
%largest sizes studied.

%This makes it hard to be absolutely sure that there is
%in fact a phase transition for the small field of $h_{r}=0.1$. Our non-linear
%Levenberg-Marquardt fitting routine gives $T_{SG} = 0.170 \pm 0.022$ from
%$\xi_{SG}$, and $T_{SG} = 0.196 \pm 0.015$ from $\chi_{SG}$.

\begin{figure}
\begin{center}
\includegraphics[width=\columnwidth]{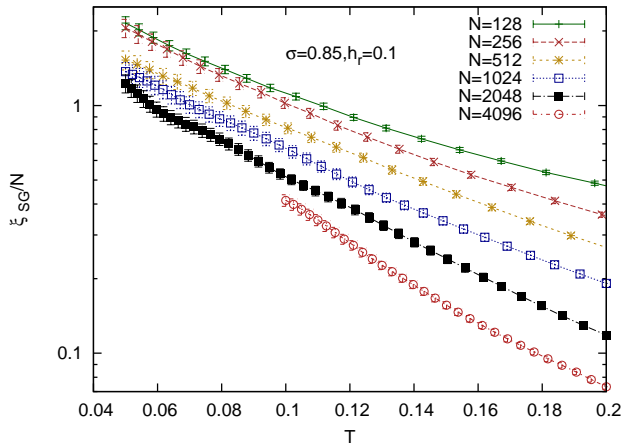}
\caption{(Color online)
A finite size scaling plot of data for $\xi_{SG}$ for $\sigma=0.85$. The magnetic
field is $h_{r}=0.1$.
No intersections are found for the range of temperatures that we simulate indicating that 
there is no phase transition in this range.
For comparison, the zero field transition temperature
is~\cite{sharma:11a} $T_c \simeq 0.165$.
}
\label{fig:corrsg_0.85_high}
\end{center}
\end{figure}

\begin{figure}
\begin{center}
\includegraphics[width=\columnwidth]{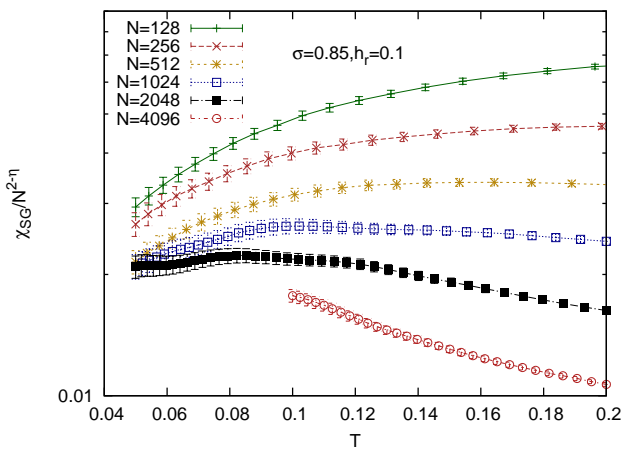}
\caption{(Color online)
A finite size scaling plot of data for $\chi_{SG}$ for $\sigma=0.85$. The magnetic
field is $h_{r}=0.1$.
At $T \simeq 0.05$, the largest sizes for which we have data, $N = 512, 1024$
and $2048$ merge together.
For comparison, the zero field transition
temperature
is~\cite{sharma:11a} $T_c \simeq 0.165$.
%No intersections are found for the range of temperatures that we simulate.
%Indicates that there is no phase transition, in agreement with the conclusion
%from $\xi_{SG}$. Compare with the zero-field transition temperature of $T_{SG}
%= 0.166 \pm 0.004$ with $\xi_{SG}$, and $T_{SG} = 0.167 \pm 0.001$ with
%$\chi_{SG}$. 
}
\label{fig:khisg_0.85_high}
\end{center}
\end{figure}

\begin{figure}
\begin{center}
\includegraphics[width=\columnwidth]{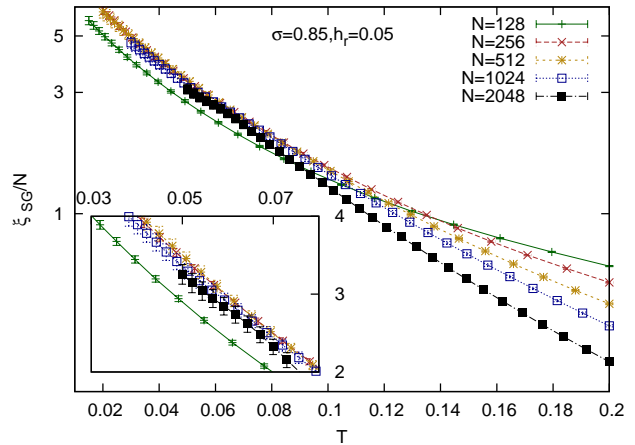}
\caption{(Color online)
A finite size scaling plot of data for $\xi_{SG}$ for $\sigma=0.85$. The magnetic
field is $h_{r}=0.05$.
%Compare with the zero-field transition temperature of $T_{SG} = 0.166 \pm 0.004$ with 
%$\xi_{SG}$, and $T_{SG} = 0.167 \pm 0.001$ with $\chi_{SG}$. 
The data appears to merge at the lowest temperatures.
The inset shows an enlargement of a region of the data.
}
\label{fig:corrsg_0.85_low}
\end{center}
\end{figure}

\begin{figure}
\begin{center}
\includegraphics[width=\columnwidth]{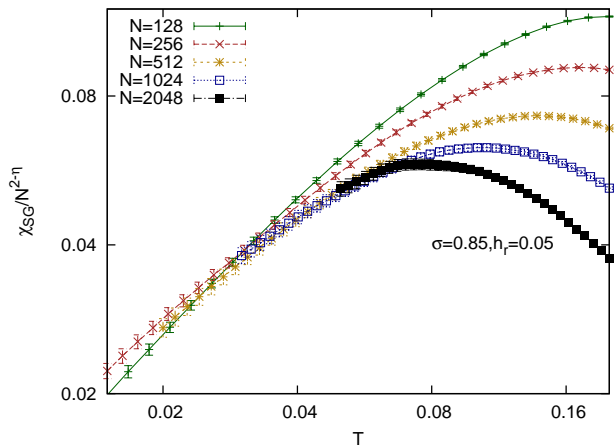}
\caption{(Color online)
A log-log
finite size scaling plot of data for $\chi_{SG}$ for $\sigma=0.85$. The magnetic
field is $h_{r}=0.05$.
%The figure shows the finite size scaling figure for $\sigma=0.85$ with
%$\chi_{SG}$, for a magnetic field of $h_{r}=0.05$. Compare with the zero-field
%transition temperature of $T_{SG} = 0.166 \pm 0.004$ with $\xi_{SG}$, and
%$T_{SG} = 0.167 \pm 0.001$ with $\chi_{SG}$. 
The data appears to merge at the lowest temperatures.
}
\label{fig:khisg_0.85_low}
\end{center}
\end{figure}

\subsection{$\sigma=0.85$}

According to Eq.~\eqref{deff}, $\sigma = 0.85$ corresponds to a short-range
model in
close to three dimensions. In their study of the Ising version of this model,
Ref.~[\onlinecite{katzgraber:09}] did not find an AT line for this value of
$\sigma$. Ref.~[\onlinecite{leuzzi:09}] did not consider this value of
$\sigma$.  

We show finite-size scaling plots for $\xi_{SG}$ and $\chi_{SG}$ for $h_r = 0.1$
and $0.05$ in Figs.~\ref{fig:corrsg_0.85_high}--\ref{fig:khisg_0.85_low}. For
$h_r = 0.1$, the $\xi_{SG}$ data does not intersect down to the lowest
temperature, 0.05, which is to be compared with the zero field transition
temperature~\cite{sharma:11a} $T_c \simeq 0.165$. The $\chi_{SG}$ data for
$h_r = 0.1$
seem to merge at the lowest temperature for the largest sizes that could be
studied in this region.

For $h_r = 0.05$, the data for \textit{both}
$\xi_{SG}$ and $\chi_{SG}$ merge at the lowest temperatures. One possible
explanation of this is that the critical field at $T=0$ is 
$h_c \lesssim 0.05$. Another
possibility is that, for this small field, we are in a crossover region
between zero-field behavior, where there is a transition, to finite-field
behavior where there is none.

According to Eq.~\eqref{hcMF}, in mean field theory the zero-temperature critical
field, $h_c$ is $1.732$ times the zero-field transition temperature $T_c$.
If we assume, based on our data, that $h_c \lesssim 0.05$ then, using $T_c
\simeq 0.165$ from Ref.~[\onlinecite{sharma:11a}],
we have a ratio $h_c / T_c \lesssim 0.30$, which is about 17\% of the
mean field result, i.e.~\textit{considerably} smaller.
Hence, if the data in
Figs.~\ref{fig:corrsg_0.85_low} and \ref{fig:khisg_0.85_low} is interpreted
to show a critical field of around 0.05, then this is \textit{much}
smaller than in
mean field theory, even allowing for the (substantial) reduction in the
zero-field $T_c$ relative to the mean field prediction.

\section{Conclusion}
\label{sec:conclusions}

We have studied existence or otherwise of the AT line in the 1-dimensional
Heisenberg spin glass with interactions which fall off as a power
of the distance.  We are able to study a large range of sizes in the
temperature range of interest: $N \le 16384$ for $\sigma = 0.6$, $N\le 4096$
for $\sigma = 0.75$, and $N \le 2048$ for $\sigma = 0.85$ (up
to 4096 at somewhat higher
temperatures).

For $\sigma = 0.6$, which is in the mean-field regime ($\sigma < 2/3$),
we find an AT line.
%consistent with earlier work~\cite{katzgraber:09,leuzzi:09} for the analogous
%Ising model.
For $\sigma = 0.75$, which is in the non-mean-field regime and
corresponds to a short-range model with dimension about 4, the data does
appear to find a phase transition for a field $h_r = 0.1$,
though the intersection temperatures in the finite-size scaling plots drop
for the largest sizes, which might give one pause to accept that conclusion
with certainty.  For $\sigma = 0.85$, which corresponds to a dimension of
about 3, the data for $h_r=0.05$ can be interpreted as indicating that this
value of field is close to a critical AT field in the limit of zero
temperature. However, it
can also be interpreted as indicating a crossover between the
zero-field transition and behavior in a field which has no transition.

For the corresponding Ising model, Ref.~[\onlinecite{katzgraber:09}] finds an
AT line in the mean-field regime but not in the non-mean-field regime. This
conclusion was challenged in Ref.~[\onlinecite{leuzzi:09}], who claim that
there is also an AT line in the non-mean-field regime, at least up for $\sigma$
between $2/3$ and $0.75$.  A motivation for the present study was to see if
a clearer numerical picture can emerge from the Heisenberg spin glass, where
it is possible to study larger sizes than for the Ising case.
Unfortunately, it seems that corrections
to scaling are quite large in the Heisenberg case, so we are not able to give
a precise value for the lower critical dimension of the AT line from the data
in our paper. If there is no AT line,  the
system breaks up into domains of size $\ell$ (Imry-Ma length) which
can be large at low temperatures. A possible explanation of our results is
therefore
that $\ell(T \to 0)$ for $\sigma = 0.75, h_r = 0.1$ is larger than
the largest system size, $N = 4096$, 
and that for $\sigma = 0.85, h_r = 0.05$, $\ell(T \to 0)$ is
about equal to the largest system size at the lowest temperature, $N = 2048$.

\acknowledgments
We acknowledge support from the NSF under Grant DMR-0906366. We also thank the
Hierarchical Systems Research Foundation for generous access to its
computers. We would like to thank Mike Moore for his comments on a draft of
this manuscript.

\bibliography{refs,comments}

\begin{thebibliography}{25}
\bibitem[{fss()}]{fss:gtlcd}
\bibinfo{note}{For a discussion of how finite-size scaling is modified in the
  region of mean-field exponents, see for example
  Refs.~[\onlinecite{brezin:82,brezin:85,binder:85,luijten:99,jones:05}].}
\end{thebibliography}

\end{document}